\documentclass[12pt]{article}
\usepackage[dvipdfmx]{graphicx}
\usepackage{mathrsfs}
\usepackage{amssymb}
\usepackage{color}
\usepackage{amsmath}
\usepackage{ascmac}
\usepackage{amsthm}
\usepackage{booktabs}
\usepackage{lscape}
\usepackage{rotating}
\usepackage{dcolumn}
\usepackage{longtable}
\usepackage{tabularx}
\usepackage{dcolumn}
\usepackage{arydshln}
\usepackage{bm}
\usepackage{url}
\usepackage{caption}
\usepackage{subfig}
\usepackage{setspace}
\usepackage{cases}
\usepackage{comment}

\usepackage[version=3]{mhchem}

\newcommand{\etal}{et al.}

\newcommand{\vx}{\mathbf{x}}

\newcommand{\vz}{\mathbf{z}}

\newcommand{\vdelta}{\bm{\delta}}

\newcommand{\vphi}{\bm{\phi}}

\makeatletter
  \renewenvironment{thebibliography}[1]{%

    \section*{\refname}\@mkboth{\refname}{\refname}%
    \list{\@biblabel{\@arabic\c@enumiv}}%
          {\settowidth\labelwidth{\@biblabel{#1}}%
          \leftmargin\labelwidth
          \advance\leftmargin\labelsep
          \@openbib@code
          \usecounter{enumiv}%
          \let\p@enumiv\@empty
          \renewcommand\theenumiv{\@arabic\c@enumiv}}%
    \sloppy
    \clubpenalty4000
    \@clubpenalty\clubpenalty
    \widowpenalty4000%
    \sfcode`\.\@m}
    {\def\@noitemerr
      {\@latex@warning{Empty `thebibliography' environment}}%
    \endlist}
\makeatother
\begin{document}

\title{
\begin{spacing}{1.0}
Particulate Air Pollution, Birth Outcomes, and Infant Mortality: Evidence from Japan's Automobile Emission Control Law of 1992
\end{spacing}
}

\author{
Tatsuki Inoue\thanks{Graduate School of Economics, The University of Tokyo, 7-3-1, Hongo, Bunkyo-ku, Tokyo
113-0033, Japan (E-mail: inoue-tatsuki245@g.ecc.u-tokyo.ac.jp).}, 
Nana Nunokawa\thanks{Department of Industrial Engineering, School of Engineering, Tokyo Institute of Technology, 2-12-1, Ookayama, Meguro-ku, Tokyo 152-8552, Japan (E-mail: nunokawa.n.aa@m.titech.ac.jp).},
Daisuke Kurisu\thanks{Department of Industrial Engineering, School of Engineering, Tokyo Institute of Technology, 2-12-1, Ookayama, Meguro-ku, Tokyo 152-8552, Japan (E-mail: kurisu.d.aa@m.titech.ac.jp).},
~and 
Kota Ogasawara\thanks{Corresponding author:~Department of Industrial Engineering, School of Engineering, Tokyo Institute of Technology, 2-12-1, Ookayama, Meguro-ku, Tokyo 152-8552, Japan (E-mail: ogasawara.k.ab@m.titech.ac.jp).}
}

\date{\today}
\maketitle
\begin{abstract}
\begin{spacing}{1.0}
This study investigates the impacts of the Automobile NO$_{\text{x}}$ Law of 1992 on ambient air pollutants and fetal and infant health outcomes in Japan.
Using panel data taken from more than 1,500 monitoring stations between 1987 and 1997, we find that NO$_{\text{x}}$ and SO$_{\text{2}}$ levels reduced by 87\% and 52\%, respectively in regulated areas following the 1992 regulation.
In addition, using a municipal-level Vital Statistics panel dataset and adopting the regression differences-in-differences method, we find that the enactment of the regulation explained most of the improvements in the fetal death rate between 1991 and 1993.
This study is the first to provide evidence on the positive impacts of this large-scale automobile regulation policy on fetal health.
\end{spacing}

\bigskip


\bigskip


\end{abstract}

\newpage
%-------------------------------------------------------------------------------
%-------------------------------------------------------------------------------
\section{Introduction} \label{sec:intro}

A growing body of the literature in the field of health economics and medical studies has found that ambient air pollution has serious health consequences.
Air pollution affects not only adult well-being, but also fetal, infant, and child health (Dockery \etal~1993; Schwartz 2004; Currie and Neidell 2005; Bateson and Schwartz 2007; Currie \etal~2009; Coneus and Spiess 2012; Janke 2014).
Moreover, recent studies have found that high levels of air pollution can also be directly linked to suicidal behavior and inversely related to happiness and life expectancy (Li \etal~2014; Ng \etal~2016; Hill \etal~2019).
The weight of this evidence has led to the consensus that air pollution control regulations are required to maintain well-being, especially in developing countries (Greenstone and Hanna 2014).

In this study, we contribute to the literature by analyzing the impacts of Japan's Automobile NO$_{\text{x}}$ Law of 1992 on birth outcomes and infant mortality.
The extensive previous literature has studied the health impacts of regulations on pollutant emissions from power plants (Luechinger 2014; Tanaka 2015), nationwide air pollution control policy (Sanders and Stoecker 2015; Lee \etal~2018), and the introduction of an emissions market for nitrogen oxides (Desch\'enes~\etal 2017).
The negative effects of air pollutants emitted from automobiles on infant health have also been highlighted (Currie~\etal~2009; Coneus and Spiess 2012).
However, the impacts of automobile emission regulations on fetal and infant health have been understudied.
While the related study by Beatty and Shimshack (2011) focused on a localized emission reduction program on school buses in the state of Washington, we evaluate a generalized emission reduction program on automobiles across Japan.
As health outcomes, we investigate and measure the impacts on several fetal and infant health outcomes in more detail than in their study.

As regulated areas, the 1992 regulation selected 196 municipalities exceeding the emission control standard in Saitama, Chiba, Tokyo, Kanagawa, Osaka, and Hyogo prefectures.
In these areas, a stringent regulation was enforced; trucks, buses, and even special motor vehicles such as ambulances were regulated through motor vehicle inspections.
We use this quasi-experimental setting induced by the regulation to compare the changes both in air pollutant concentrations and in birth and infant health outcomes between regulated and non-regulated areas before and after the treatment, using the differences-in-differences (DID) methodology.

To conduct the analysis, we first use panel data taken from more than 1,500 monitoring stations between 1987 and 1997.
As air quality outcomes, we consider a wide variety of pollutants, namely nitrogen dioxide (NO$_{\text{X}}$), sulfur dioxide (SO$_{\text{2}}$), photochemical oxidants (O$_\text{x}$), and suspended particulate matter (SPM).
Although related studies have compared air quality in regulated and non-regulated areas, we assume that the treatment effects vary between the core regulated areas and surrounding areas (Greenstone 2004; Auffhammer and Kellogg 2011). 
Using these monitoring data with location information, we thus investigate the neighboring effects of the regulation by controlling for the meteorological features around monitoring stations.
We then draw on municipal-level data on fetal and infant health outcomes from the \textit{Vital Statistics of Japan} that contain \textit{all} the birth and death records of the six target prefectures in 1991 and 1993.
The fetal death rate (FDR), low-birth weight rate (LBWR), infant mortality rate (IMR), and neonatal mortality rate (NMR) are used to comprehensively analyze the potential effects of the regulation.

We find that the regulation significantly improved air quality in regulated areas.
Our estimates suggest that NO$_{\text{x}}$ and SO$_{\text{2}}$ levels reduced by 87\% and 52\%, respectively in regulated areas following the 1992 regulation.
Accordingly, we find that the enactment of the regulation law reduced the risk of fetal deaths.
Our estimate suggests that the regulation reduced the FDR by $3.5$‰, fully explaining the improvements in the rates in regulation areas between 1991 and 1993.

The remainder of this paper is structured as follows.
Section~\ref{sec:back} provides a brief overview of the historical background of the regulation policies.
Section~\ref{sec:data} describes the data used.
Section~\ref{sec:analysis} presents our empirical strategy and the results.
Section~\ref{sec:con} concludes.

%-------------------------------------------------------------------------------
%-------------------------------------------------------------------------------
\section{Background} \label{sec:back}

The occurrence of diseases due to environmental pollution emerged as a serious problem in Japan during the country's period of rapid economic growth in 1954--1973 (Harada 1995).
This led to the enactment of the Basic Act for Environmental Pollution Control in 1967.
This law established environmental quality standards on air pollutants.
To ensure compliance with these standards, the Air Pollution Control Act, which aimed to limit emissions of air pollutants from factories, business establishments, and automobiles, was then enacted in 1968.
However, NO$_\text{2}$ concentration failed to meet this standard---even by 1985; Indeed, it subsequently exhibited a worsening trend because of the increase in the use of diesel automobiles (Air Quality Bureau of the Environment Agency and Automobile NO$_\text{x}$ Law Research Group 1994).

Thus, to reduce NO$_\text{x}$ emissions, especially NO$_\text{2}$, from automobiles, the Automobile NO$_\text{x}$ Law was enacted in 1992.
This law selected 196 municipalities in Saitama, Chiba, Tokyo, Kanagawa, Osaka, and Hyogo prefectures as regulated areas based on two considerations: traffic flow in these areas was heavy and it would be difficult for them to comply with the environmental standards for NO$_\text{2}$ using only existing regulations.
In these areas, trucks, buses, and special motor vehicles such as ambulances began to be regulated through motor vehicle inspections if they did not satisfy the emission control standard.
As a result of the regulation, the use of diesel automobiles that emit considerable amounts of toxic exhaust gas was controlled (\ref{sec:appa}).

%-----
%Figure 
\begin{figure}[t]
    \centering
    \subfloat[NO$_\text{x}$] {\label{fig:NOx}\includegraphics[width=0.5\textwidth]{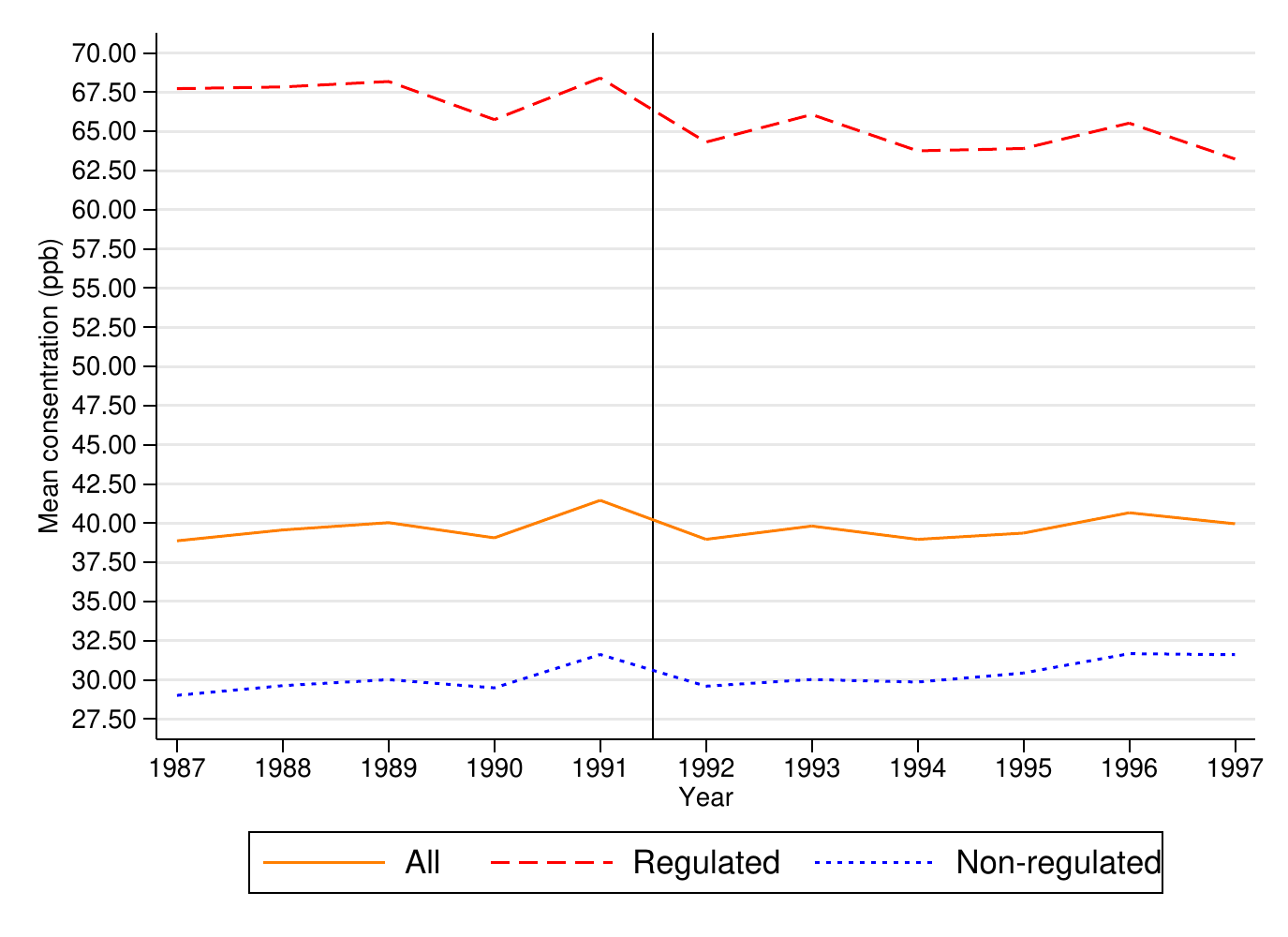}}
    \subfloat[SO$_\text{2}$]{\label{fig:SO2}\includegraphics[width=0.5\textwidth]{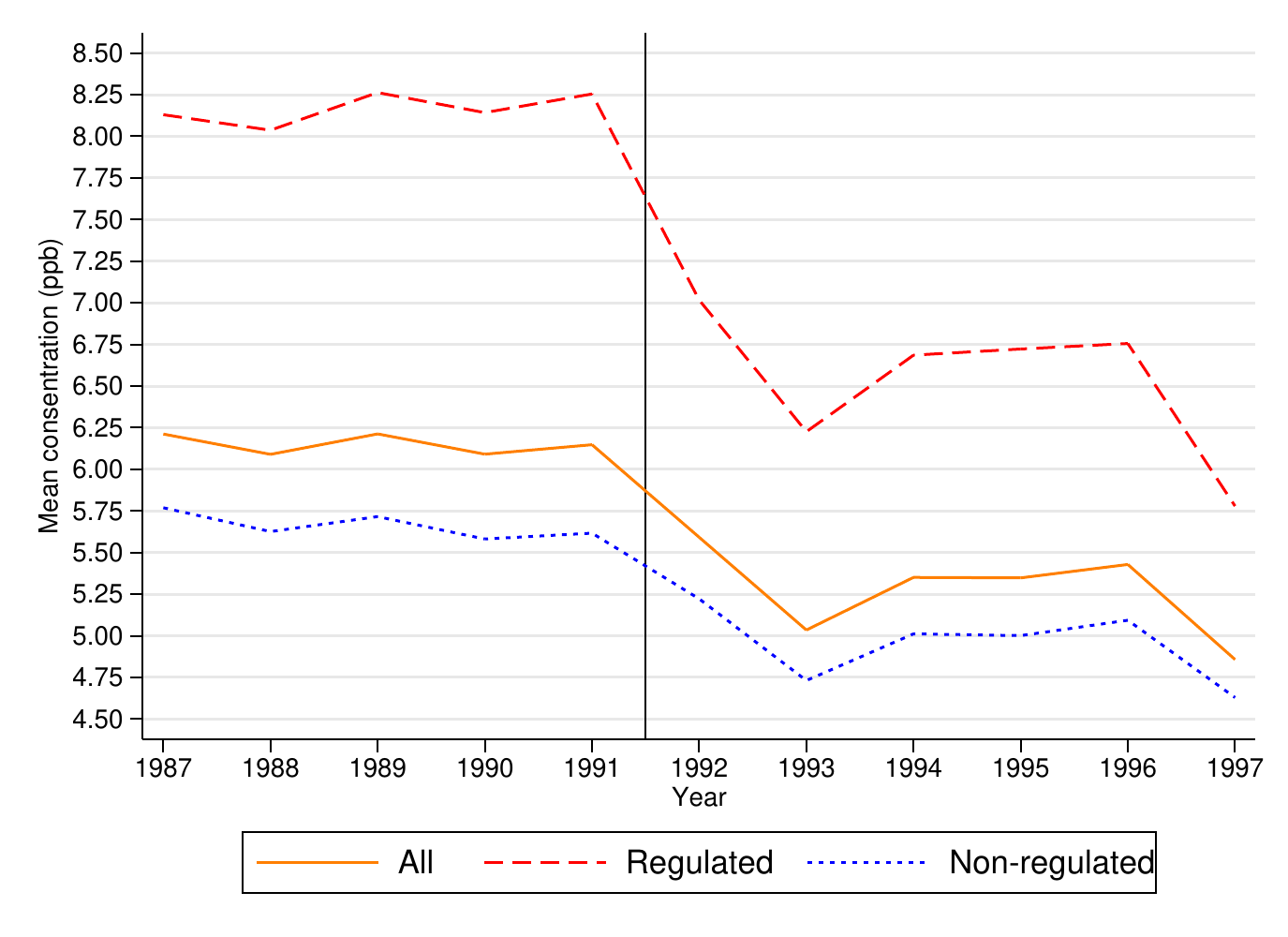}}
    \caption{Annual average concentrations of NO$_\text{x}$ and SO$_\text{2}$, 1987--1997}
    \label{fig:NOxSO2}
    \scriptsize{\begin{minipage}{160mm}
    Notes:
    The solid, dashed, and dotted lines indicate average concentrations of NO$_\text{x}$ and SO$_\text{2}$ in all areas, regulated areas, and non-regulated areas, respectively.
    The vertical line indicates the timing of the enactment of the Automobile NO$_\text{x}$ Law.
    Sources: National Institute for Environmental Studies, Environmental Database (Appendix~\ref{sec:appb1}).
    \end{minipage}}
\end{figure}
%------

Figure~\ref{fig:NOx} displays the annual average concentrations of NO$_\text{x}$ and SO$_\text{2}$ from 1987 to 1997, showing that the level of NO$_\text{x}$ in regulated areas declined when the regulation was introduced.
The mean NO$_{\text{x}}$ concentrations of regulated areas before and after the enactment of the regulation was 67.60 and 63.77 parts per billion (ppb), respectively.
The level of SO$_\text{2}$ also declined between 1991 and 1993, as shown in Figure~\ref{fig:SO2}.
The decrease in SO$_\text{2}$ level in regulated areas was larger than that in non-regulated areas.
Indeed, no such greater declines in SO$_\text{2}$ than NO$_\text{x}$ had been highlighted in previous related studies (Arimura and Iwata 2008; Wakamatsu \etal~2013).
However, diesel exhaust emissions include not only NO$_\text{x}$ but also SO$_\text{2}$, because the sulfur contained in light gas oil is oxidized by combustion (\ce{S + O2 -> SO2}).
Hence, it is clear that the pronounced SO$_\text{2}$ decline can be attributed to the exhaust emission regulation, highlighting the need to reassess the effectiveness of the Automobile NO$_\text{x}$ Law.

%-------------------------------------------------------------------------------
%-------------------------------------------------------------------------------
\section{Data} \label{sec:data}

\subsection{Air pollutant concentrations} \label{sec:dataap}

%-------------------------------------
%Figure
\begin{figure}[t]
\centering
\includegraphics[width=\textwidth]{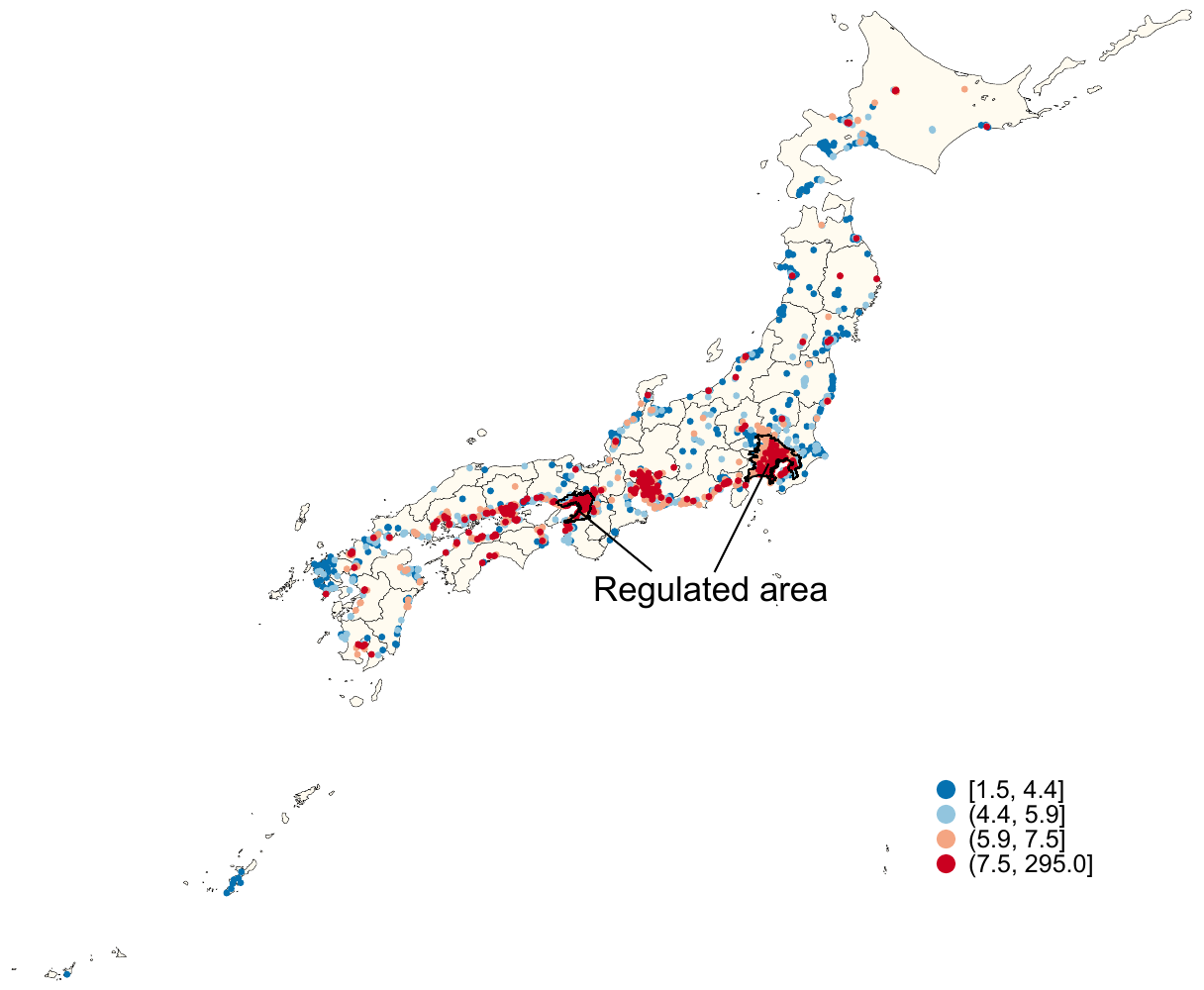}
\caption{Spatial distribution of SO$_\text{2}$ concentration and regulated areas}
\label{fig:map}
{\scriptsize
\begin{minipage}{160mm}
Notes: The SO$_{\text{2}}$ concentrations (ppb) represent the mean value during 1987--1991.
The data on longitudes and latitudes are available for only those stations still in use in 2001, and this figure covers 91.4\% ($1,519/1,662$) of the stations used in the regression analyses for SO$_\text{2}$ concentration (NO$_\text{x}$ shows a similar distribution).
Sources: National Institute for Environmental Studies, Environmental Database (Appendix~\ref{sec:appb1}).
\end{minipage}
}
\end{figure}
%-------------------------------------

For our empirical analyses, we use a panel dataset on the annual average air pollutant concentrations of NO$_{\text{X}}$, SO$_{\text{2}}$, O$_\text{x}$, and SPM at the monitoring station level between 1987 and 1997 taken from the \textit{Environmental Database} provided by the National Institute for Environmental Studies (units in ppb).
In this study, O$_\text{x}$ and SPM are also considered to be secondary pollutants because they can be emitted by diesel automobiles.
Figure~\ref{fig:map} illustrates the location of regulated areas and the stations that monitored SO$_{\text{2}}$ concentration.
There were more than 1,500 stations nationally around 1992.
Regulated areas had as many as 342 stations, or 20.6\% of the total number.

\subsection{Birth outcomes and infant mortality} \label{sec:datahealth}

For fetal and infant health outcomes, we constructed a municipal-level panel dataset from 435 municipalities in 1991 and 1993 using the \textit{Vital Statistics of Japan}.
As Sanders and Stoecker (2015) discussed, birth- and death-related outcome variables such as the FDR and IMR calculated from survey samples are more likely to suffer sample selection issues.
However, our comprehensive registration-based birth and death records enable us to overcome such potential selection issues in the regression analyses. 
As noted earlier, we adopt the FDR, LBWR, IMR, and NMR in our statistical analyses.
The FDR is the number of fetal deaths per 1,000 births, the LBWR is defined as the number of births with a weight less than 2,500 g per 100 births, the IMR is the number of infant deaths (within a year of birth) per 100 live births, and the NMR is the number of neonatal deaths (within four weeks of birth).

\subsection{Control variables} \label{sec:datacon}

We use meteorological data as the control variables in the air pollutant regressions because weather shocks can affect air quality (Silva et al.~2017); strong wind and rain may carry away air pollutants and short periods of sunshine may mitigate ozone production.
If meteorological conditions changed at the same time and in the same place as the enforcement of the Automobile NO$_{\text{x}}$ Law, we would fail to capture the regulation's effects.
To deal with this potential issue, we use variables controlling for the number of days with rainfall of over 1 mm, the number of days with a maximum wind speed of over 10 m/s, and the percentage of sunshine hours in a year.
Each air pollution monitoring station uses meteorological data from the nearest weather monitoring station.

In the health regressions, we use data on public health and regional standards of living.
We consider the coverage of hospitals because accessibility to medical institutions may be negatively correlated with health outcomes.
The share of low-income households, namely the number of households receiving welfare benefits per 100 households, is also considered.
The Japanese economy was already developed in the 1990s, and thus the density of low-income households is a more suitable control variable than average household income for capturing adverse birth and infant health outcomes (Cabinet Office 2016).

\ref{sec:appb} reports the details and summary statistics of the dependent and control variables.

%-------------------------------------------------------------------------------
%-------------------------------------------------------------------------------
\section{Empirical Analysis} \label{sec:analysis}

\subsection{Air pollutant concentrations} \label{sec:ap}

We begin our empirical analysis by examining the effects of the 1992 regulation on air pollutant concentrations.
To do so, we employ a fixed effects model in the spirit of the DID approach in the following form:
\begin{eqnarray}
\textit{Pollutant}_{it} = \alpha + \beta \textit{Regulation}_{i} \times \textit{Post}_{t} + \gamma \textit{Neighborhood}_{i} \times \textit{Post}_{t} + \vx'_{it} \vdelta + \nu_{i} + \mu_{t}  + \epsilon_{it} \label{eq:base}
\end{eqnarray}
where $i$ indexes monitoring stations and $t$ indexes years.
The dependent variable $\textit{Pollutant}_{it}$ represents average NO$_\text{x}$, SO$_\text{2}$, O$_\text{x}$, or SPM concentrations (ppb).
$\textit{Regulation}_{i}$ represents an indicator variable for regulated areas and $\textit{Post}_{t}$ is an indicator variable for the post-regulation period.
Therefore, our parameter of interest is $\beta$, and its estimate $\hat{\beta}$ can be interpreted as a potential effect of the regulation.
To capture the external effects, we also include the interaction term $\textit{Neighborhood}_{i} \times \textit{Post}_{t}$, where $\textit{Neighborhood}_{i}$ is an indicator variable coded one if $i$ is not included in the regulated areas but is in the six included prefectures.
If the estimate $\hat{\gamma}$ is statistically significantly negative, therefore, the regulation might have had positive neighboring effects.
$\vx_{it}$ is a vector of the meteorological control variables introduced in Section~\ref{sec:datacon}.
$\nu_{i}$ and $\mu_{t}$ represent the monitoring station and year fixed effects, respectively.
$\epsilon_{it}$ is a random error term.

%------------------------------------
%Table 
\begin{table}
\begin{center}
\caption{Effects of the Automobile NO$_\text{x}$ Law on air pollutants}
\label{tab:result}
\footnotesize

\begin{tabular*}{160mm}{l@{\extracolsep{\fill}}cccc}
\toprule

&(1) NO$_\text{x}$&(2) SO$_\text{2}$&(3) O$_\text{x}$&(4) SPM\\
&[67.86]&[8.2]&[22.74]&[49.86]\\\hline

\textit{Regulation} $\times$ \textit{Post}
&$-$3.458***&$-$0.893***&$-$1.100***&$-$1.450***\\
&(0.527)&(0.102)&(0.230)&(0.362)\\
\textit{Neighborhood} $\times$ \textit{Post}
&$-$0.653&0.001&$-$0.018&0.263\\
&(0.627)&(0.130)&(0.444)&(0.517)\\\hline

Meteorological controls	&Yes&Yes&Yes&Yes\\
Station fixed effects	&Yes&Yes&Yes&Yes\\
Year fixed effects		&Yes&Yes&Yes&Yes\\

Observations
&14,085&13,743&9,286&12,193\\
Clusters
&\, 1,312&\, 1,274&\, \ 860&\, 1,186\\
Adj. $R$-squared&0.4516&0.5425&0.4322&0.3535\\
\bottomrule
\end{tabular*}

{\scriptsize
\begin{minipage}{160mm}
Notes: Each dependent variable represents annual average concentration (ppb). The mean values of the dependent variables in regulated areas in the pre-intervention period are reported in brackets.
The meteorological control variables include the number of days with rainfall of over 1 mm, the number of days with a maximum wind speed of over 10 m/s, and the percentage of sunshine hours in a year.
***, **, and * represent statistical significance at the 1\%, 5\%, and 10\% levels, respectively.
Standard errors clustered at the monitoring station level are in parentheses.
\end{minipage}
}
\end{center}
\end{table}
%------------------------------------

Table~\ref{tab:result} presents the results.
As shown in Columns~(1)--(4), the estimated effects of the regulation are statistically significantly negative across all specifications.
By contrast, the estimated coefficients of \textit{Neighborhood} $\times$ \textit{Post} are close to zero and statistically insignificant.
Moreover, our results are largely unchanged if we exclude this neighborhood interaction from Eq.~\ref{eq:base} (Appendix~\ref{sec:appc1}).
This finding suggests that the regulation did not particularly affect pollutants in the neighborhood of regulated areas.

The estimates reported in Columns~(1) and (2) indicate that the law decreased the levels of NO$_\text{x}$ and SO$_\text{2}$ in regulated areas by roughly 5\% ($3.458/67.86$) and 11\% ($0.893/8.2$), respectively.
Clearly, the 1992 regulation significantly reduced not only NO$_\text{x}$ but also SO$_\text{2}$ pollution, as discussed in Section~\ref{sec:back}.
Columns~(3) and (4) confirm that the regulation also reduced the O$_\text{x}$ and SPM concentrations; the decreases in O$_\text{x}$ and SPM attributed to the regulation are calculated as roughly 5\% ($1.1/22.74$) and 3\% ($1.45/49.86$), respectively.

\subsection{Birth outcomes and infant mortality} \label{sec:health}

%------------------------------------
%Table 
\begin{table}
\begin{center}
\caption{Effects of the Automobile NO$_\text{x}$ Law on fetal and infant health outcomes}
\label{tab:did}
\footnotesize

\begin{tabular*}{160mm}{l@{\extracolsep{\fill}}cccc}
\toprule

		&(1) FDR	&(2) LBWR	&(3) IMR	&(4) NMR	\\
		&[37.24]	&[6.62]		&[4.34]		&[2.27]	\\\hline
\textit{Regulation}
&1.572&0.180&1.646&0.886\\
&(4.799)&(0.449)&(1.590)&(0.881)\\
\textit{Post}
&0.347&0.626**&$-$0.010&$-$0.003\\
&(1.576)&(0.250)&(0.558)&(0.407)\\
\textit{Regulation} $\times$ \textit{Post}
&$-$3.534**&$-$0.287&0.192&0.156\\
&(1.668)&(0.259)&(0.593)&(0.428)\\\hline

Control variables
&Yes&Yes&Yes&Yes\\
City-county FE
&Yes&Yes&Yes&Yes\\
Observations
&870&870&870&870\\
Municipalities
&435&435&435&435\\
Adj. $R$-squared
&0.4352&0.1777&0.0338&0.0191\\
\bottomrule
\end{tabular*}

{\scriptsize
\begin{minipage}{160mm}
Notes: FDR, LBWR, IMR, and NMR represent the fetal death rate, low-birth weight rate, infant mortality rate, and neonatal mortality rate, respectively.
The regression shown in Column (1) is weighted by the number of births, whereas the regressions shown in Columns (2)--(4) are weighted by the number of live births.
The mean values of the outcome variables in regulated areas in the pre-intervention period are reported in brackets.
The control variables include the coverage of hospitals and proportion of households receiving welfare benefits.
***, **, and * represent statistical significance at the 1\%, 5\%, and 10\% levels, respectively.
Standard errors clustered at the city-county level are in parentheses.
The number of city-county clusters is 245.
\end{minipage}
}
\end{center}
\end{table}
%------------------------------------

Our finding in Section~\ref{sec:ap} confirms two important features of the regulation: (1) it played an important role in reducing air pollutant concentrations and (2) it did not significantly improve the effects in neighboring areas.
Considering these features, we thus investigate the potential improving effects of the regulation on a wide variety of fetal and infant health outcomes using the regression DID models in the following form:
\begin{eqnarray}
\textit{Health}_{jt} = \pi + \kappa \textit{Regulation}_{j} + \eta \textit{Post}_{t} + \theta \textit{Regulation}_{j} \times \textit{Post}_{t} + \vz'_{jt} \textbf{$\vphi$} + \lambda_{g_{j}} + u_{jt} \label{eq:did1}
\end{eqnarray}
where $j$ denotes municipalities and $t$ indexes years.
The dependent variable $\textit{Health}_{jt}$ is the FDR, LBWR, IMR, or NMR.
$\vz_{jt}$ is a vector of the control variables introduced in Section~\ref{sec:datacon}.
$\lambda_{g_{j}}$ is a city-county fixed effect, which captures the unobservable time-constant factors varying over cities and counties, and $v_{jt}$ is a random error term.
Our parameter of interest is $\theta$, and its estimate $\hat{\theta}$ can be interpreted as a potential treatment effect of the regulation on health outcomes.

Table~\ref{tab:did} presents the results.
Column (1) indicates that the estimated treatment effect on the FDR is negative and statistically significant.
This estimate suggests that the regulation reduced the FDR by $3.5$‰, accounting for roughly 10\% ($3.5/37.24$) of the mean value of the rate in regulated areas in the pre-intervention period.
The improving effect of the air pollution control act on fetal deaths is indeed consistent with the findings of previous studies (Sanders and Stoecker 2015).
Column (2) shows that the estimated effect on the LBWR is insignificant but negative.
However, we find no significant improving effects on infant and neonatal mortality rates, as shown in Columns (3) and (4).
The poor model fitting reflected in the $R$-squared values for these dependent variables (i.e., 0.0338 and 0.0191) is consistent with the fact that standards of infant health were sufficiently high in 1990s Japan.

Moreover, our results are unchanged if we cluster the standard errors at the prefecture level rather than the city-county level to take the potential correlations among cities and counties into account (Appendix~\ref{sec:appc2}).

%-------------------------------------------------------------------------------
%-------------------------------------------------------------------------------
\section{Conclusion}\label{sec:con}

The present study assessed the Automobile NO$_\text{x}$ Law in Japan using quantitative methods.
Our estimates indicate that the regulation decreased annual average NO$_\text{x}$ and SO$_\text{2}$ concentrations by 3.46 and 0.89 ppb in regulated areas, respectively.
Since the mean differences in NO$_\text{x}$ and SO$_\text{2}$ concentrations before and after the regulation are 3.97 and 1.71 (Appendix~\ref{sec:appb1}), the decreases due to the regulation accounted for approximately 87\% and 52\% of the total decreases in concentrations in regulated areas, respectively.
We also found that the regulation reduced the FDR by $3.5$‰ in regulated areas.
Since the mean difference in the FDR before and after the regulation was 2.4‰ (Appendix~\ref{sec:appb2}), the enactment of 
the regulation fully explains the improvements in the FDR between 1991 and 1993.

We contribute to the related literature in the following two ways.
First, as discussed in the Introduction, this study is the first to provide empirical evidence on the positive impacts of this large-scale automobile regulation policy.
While Beatty and Shimshack (2011) investigated a localized emission reduction program on school buses, we evaluate a more generalized emission reduction program on automobiles.
Moreover, we investigate the impacts of the regulation on several fetal and infant health outcomes not considered in their study.

Second, we provide a case study of Japan.
Previous evidence of the relationship between air pollution and infant and child health is predominantly from developed countries such as England, Germany, Sweden, and the United States (Chay and Greenstone 2003; Currie \etal 2009; Coneus and Spiess 2012; Janke 2014; Simeonova \etal~2019).
The vast majority of studies investigating the role of air quality regulations have also focused on these western developed countries (Greenstone 2004; Luechinger 2014).
In addition to the extensive studies of the case of China (Tanaka 2015; Henneman \etal~2017), recent works have started to examine the case of South Korea (Altindag \etal~2017; Lee \etal~2018).

However, the effectiveness of Japan's regulation policy remains understudied.
Despite its rapid post-war economic growth, Japan overcame its air pollution problems to a great degree by the 1980s (Wakamatsu \etal~2013) by implementing the strictest regulations in the world (Air Quality Bureau of the Environment Agency and Automobile NO$_\text{x}$ Law Research Group 1994).
In this respect, Japan is considered to be the ideal example of an Asian country for studying effective regulation policies.
Our findings support the evidence that a stringent (i.e., wide-scale inspection-based) automobile regulation can reduce the risk of fetal deaths in highly polluted areas.

%-------------------------------------------------------------------------------
%-------------------------------------------------------------------------------
\section*{Acknowledgment}\label{sec:ack}
We wish to thank the participants of the Tokyo Tech seminars for their helpful comments.
This work was supported by the Grant-in-Aid for JSPS Fellows (Grant Number: 17J03825) and JSPS KAKENHI (Grant Numbers 18H05679 and 19K13754). There are no conflicts of interest to declare. All errors are our own.

%-------------------------------------------------------------------------------
% references
%-------------------------------------------------------------------------------

%-------------------------------------------------------------------------------
% Appendix
%-------------------------------------------------------------------------------
\clearpage
\thispagestyle{empty}

\begin{center}
\qquad

\qquad

\qquad

\qquad

\qquad

\qquad

{\Large \textbf{
Online Appendices: For online publication only (Supplemental materials for review)
}}
\end{center}

%-------------------------------------------------------------------------------
% Appendix A
%-------------------------------------------------------------------------------
\clearpage
\appendix
\def\thesection{Appendix~\Alph{section}}
\def\thesubsection{\Alph{section}.\arabic{subsection}}
\setcounter{page}{1}

\section{Background appendix}\label{sec:appa}
\setcounter{figure}{0} \renewcommand{\thefigure}{A.\arabic{figure}}
\setcounter{table}{0} \renewcommand{\thetable}{A.\arabic{table}}

%------------------------------------
%Table 
\begin{table}[t]
\begin{center}
\caption{Emission control standards for automobiles}
\label{tab:stan}
\footnotesize
\begin{tabular*}{160mm}{@{\extracolsep{\fill}}rllrrllrr}
\toprule
&\multicolumn{4}{c}{Diesel}&\multicolumn{4}{c}{Gasoline}\\
\cmidrule(rl){2-5}
\cmidrule(rl){6-9}
\multicolumn{1}{l}{Weight class}&\multicolumn{1}{c}{Test mode}&\multicolumn{1}{c}{Measure}&\multicolumn{1}{c}{Max}&\multicolumn{1}{c}{Mean}&\multicolumn{1}{c}{Test mode}&\multicolumn{1}{c}{Measure}&\multicolumn{1}{c}{Max}&\multicolumn{1}{c}{Mean}\\
\hline

$\leq$1.7
&10$\cdot$15&g/km&0.48&0.25&10$\cdot$15&g/km&0.48&0.25\\
&D6&ppm&100&70&-&-&-&-\\
$1.7 < x \leq 2.5$
&10$\cdot$15&g/km&0.98&0.70&10$\cdot$15&g/km&0.98&0.70\\
&D6&ppm&210&150&-&-&-&-\\
$2.5 < x \leq 5.0$
&D13&g/kWh&6.90&5.10&13&g/kWh&6.90&5.10\\
&D6&ppm&350&260&6&ppm&600&450\\
$>$5.0
&D13&g/kWh&9.40&7.20&13&g/kWh&9.40&7.20\\
&D6&ppm&520&400&6&ppm&900&690\\

\bottomrule
\end{tabular*}

{\scriptsize
\begin{minipage}{160mm}
Notes:
The unit of weight is metric tons.
The initial character D in the second column indicates the mode for diesel automobiles.
This emission control standard was partly updated in 1999.
Sources: Air Quality Bureau of the Environment Agency and Automobile NO$_\text{x}$ Law Research Group (1994), p. 85.
\end{minipage}
}
\end{center}
\end{table}
%------------------------------------
%-----
%Figure 
\begin{figure}[t]
    \centering
    \subfloat[Number of diesel automobiles] {\label{fig:Numdie}\includegraphics[width=0.5\textwidth]{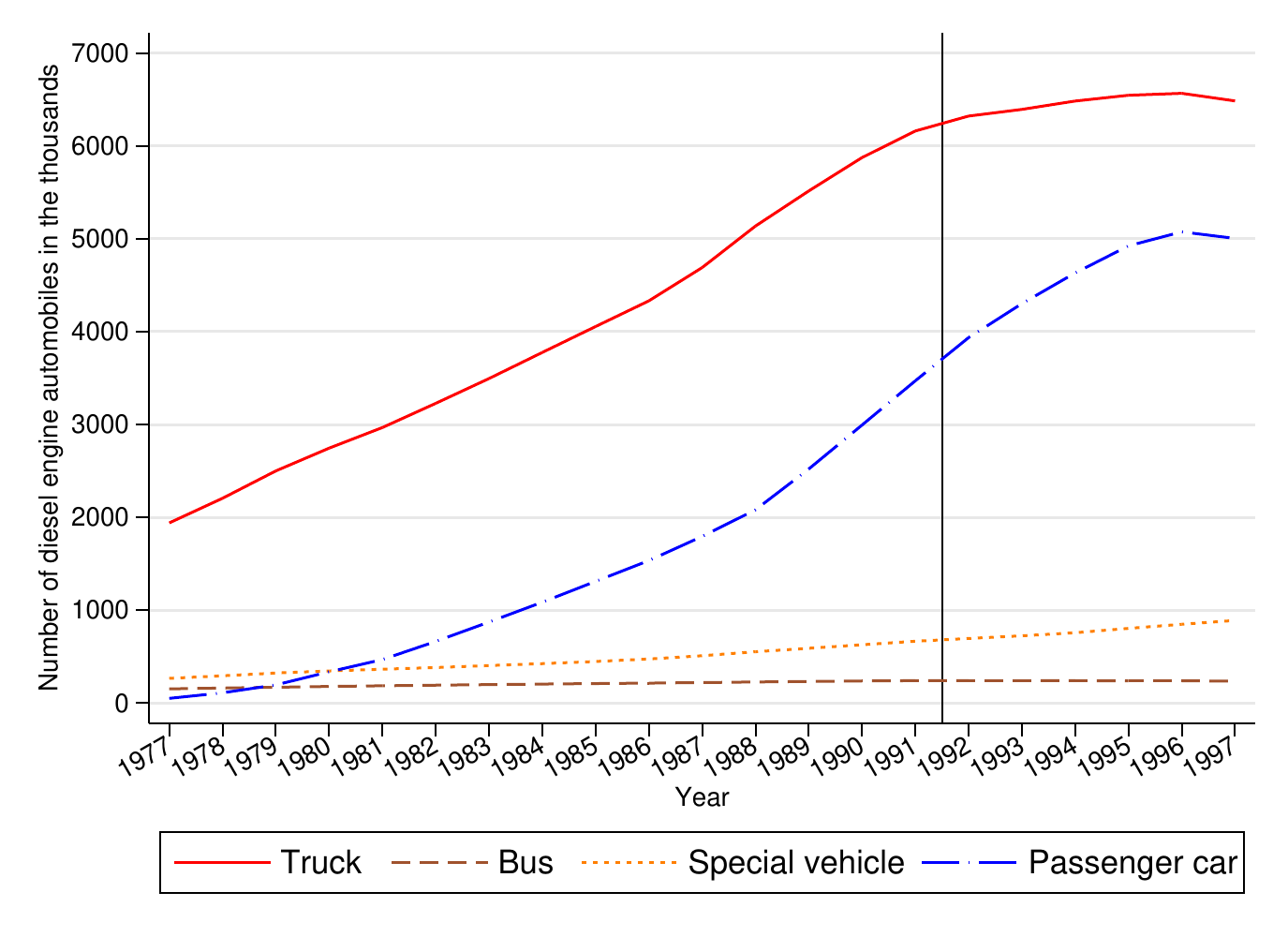}}
    \subfloat[Proportion of diesel automobiles]{\label{fig:Prodie}\includegraphics[width=0.5\textwidth]{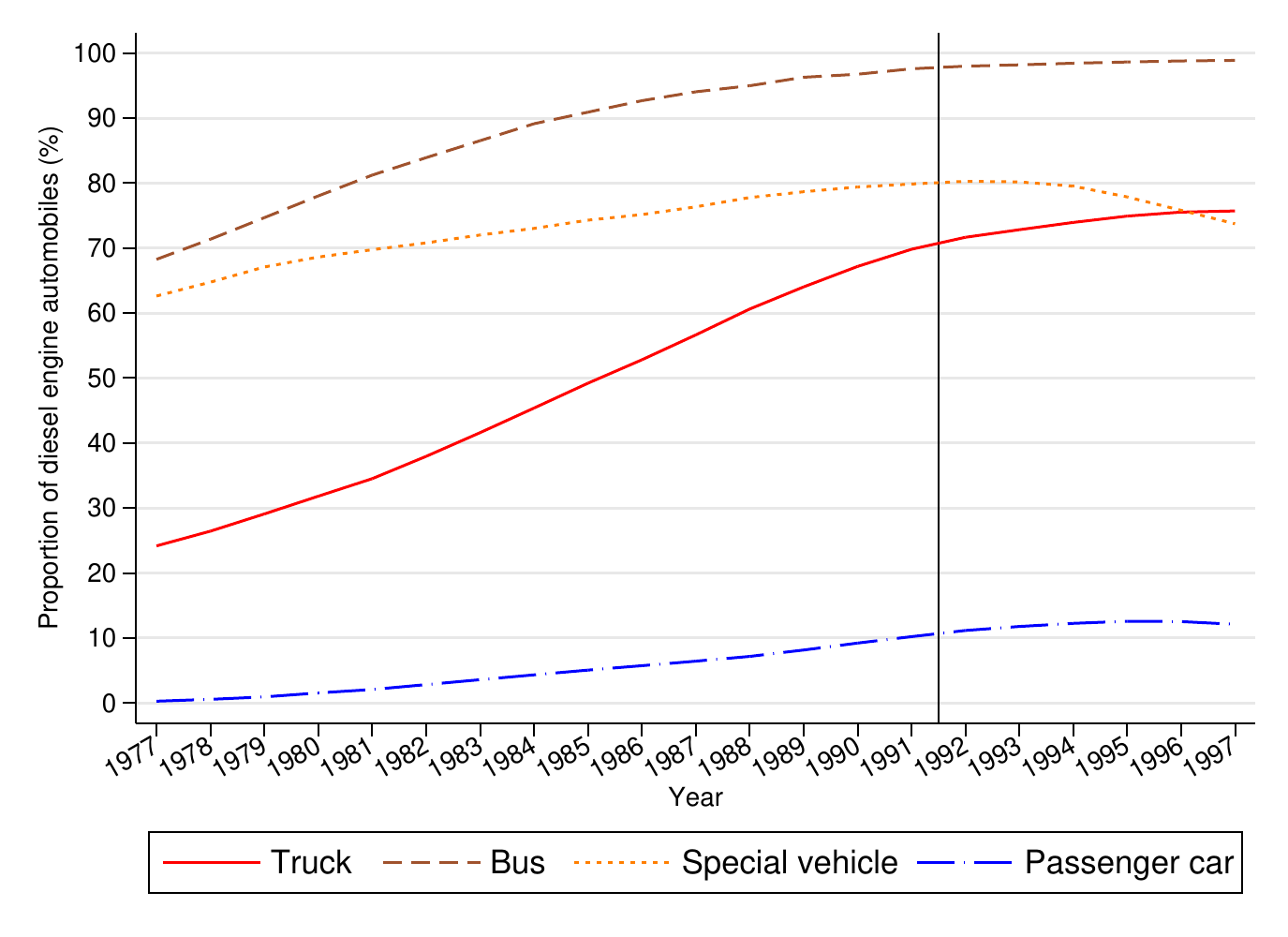}}
    \caption{Number and proportion of diesel automobiles, 1977--1997}
    \label{fig:diesel}
    \scriptsize{
    \begin{minipage}{160mm}
    Notes:
    The vertical line indicates the timing of the enactment of the Automobile NO$_\text{x}$ Law.
    Sources: Automobile Inspection \& Registration Association (1978--1998).
    \end{minipage}
    }
\end{figure}
%------

The Automobile NO$_\text{x}$ Law was enacted in 1992 to reduce NO$_\text{x}$ emissions, especially NO$_\text{2}$, from automobiles.
In regulated areas, trucks, buses, and special motor vehicles such as ambulances were regulated through motor vehicle inspections if they did not satisfy the emission control standard summarized in Table~\ref{tab:stan}.
Figure~\ref{fig:diesel} shows the numbers and proportions of diesel automobiles by vehicle classification between 1977 and 1997.
As indicated in Figure \ref{fig:Numdie}, the regulation curbed the increase in the number of diesel trucks, whereas passenger cars with diesel engines continued to increase at a constant rate around 1992.
Figure~\ref{fig:Prodie} also indicates the regulation's effect on diesel automobiles.
The increase in the proportion of diesel trucks and buses slowed when the law was enacted.
Moreover, the proportion of diesel automobiles in the special motor vehicles category decreased after the enactment of the regulation.
Although data are available only at the national level, it seems reasonable to assume that such changes were observed in regulated areas, too.

\section{Data appendix}\label{sec:appb}
\setcounter{figure}{0} \renewcommand{\thefigure}{A.\arabic{figure}}
\setcounter{table}{0} \renewcommand{\thetable}{B.\arabic{table}}

\subsection{Air pollutant concentrations}\label{sec:appb1}

%-------------------------------------
%Figure
\begin{figure}[!t]
\centering
\includegraphics[width=\textwidth]{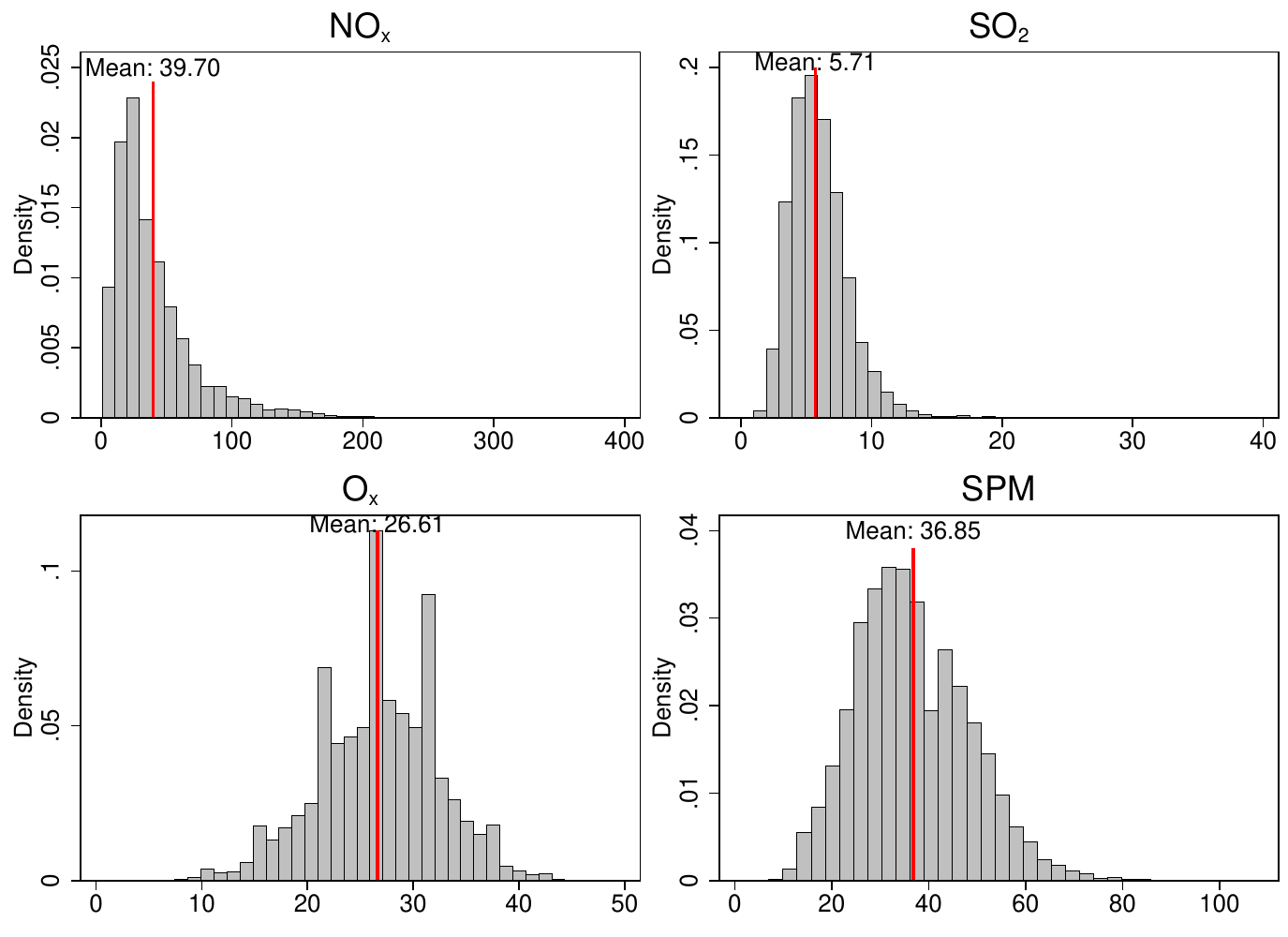}
\caption{Distribution of air pollutant concentrations}
\label{fig:hist}
{\scriptsize
\begin{minipage}{160mm}
Notes: 
Annual average concentrations (ppb) are used.
The vertical line indicates the mean value.
Sources: National Institute for Environmental Studies, Environmental Database.
\end{minipage}
}
\end{figure}
%-------------------------------------

As described in Section~\ref{sec:data}, we use data on annual average air pollutant concentrations obtained from the \textit{Environmental Database} provided by the National Institute for Environmental Studies (\url{http://www.nies.go.jp/index-e.html}, accessed September 9, 2018).
NO$_{\text{X}}$ and SO$_{\text{2}}$ are used as our primary dependent variables.
The World Health Organization (1996) provided an example of the composition of light-duty diesel engine exhaust fumes; the components are carbon dioxide, water vapor, oxygen, nitrogen, carbon monoxide, hydrocarbons, nitrogen oxides, hydrogen, sulfur dioxide, sulfates, aldehydes, ammonia, and particulates.
Thus, we also consider O$_\text{x}$ and SPM as secondary pollutants.
The mean concentrations of NO$_{\text{x}}$, SO$_{\text{2}}$, O$_{\text{x}}$, and SPM in regulated areas before 1992 were 67.601, 8.165, 22.743, and 49.792 ppb, respectively.
Figure~\ref{fig:hist} presents the histogram for air pollutant concentrations, which demonstrates that NO$_{\text{x}}$ and SO$_{\text{2}}$ levels are heavily right-skewed.
Their long tails imply that air pollution was intense in specific areas of Japan at the end of the 20th century.
Table~\ref{tab:sumap} presents the summary statistics of these pollutants by area and period.

%------------------------------------
%Table 
\begin{table}[!t]
\begin{center}
\caption{Summary statistics of the air pollutant variables}
\label{tab:sumap}
\footnotesize

\begin{tabular*}{160mm}{l@{\extracolsep{\fill}}rrrrrr}
\toprule
&\multicolumn{3}{c}{Pre-treatment}&\multicolumn{3}{c}{Post-treatment}\\
\cmidrule(rl){2-4}
\cmidrule(rl){5-7}
&\multicolumn{1}{c}{Mean}&\multicolumn{1}{c}{SD}&\multicolumn{1}{c}{N}&\multicolumn{1}{c}{Mean}&\multicolumn{1}{c}{SD}&\multicolumn{1}{c}{N}\\
\hline

\textit{Regulated areas}
&&&&&&\\
\hspace{10pt}NO$_{\text{x}}$
&67.86&40.25&2,194&63.89&34.92&2,682\\
\hspace{10pt}SO$_{\text{2}}$
&8.20&2.66&1,637&6.49&1.75&1,940\\
\hspace{10pt}O$_{\text{x}}$
&22.74&5.03&1,439&25.20&4.37&1,735\\
\hspace{10pt}SPM
&49.86&10.48&1,697&46.51&9.57&2,298\\
\\
\textit{Non-regulated areas}
&&&&&&\\
\hspace{10pt}NO$_{\text{x}}$
&29.88&23.78&5,933&30.28&22.91&7,270\\
\hspace{10pt}SO$_{\text{2}}$
&5.70&2.53&6,399&4.99&1.92&7,702\\
\hspace{10pt}O$_{\text{x}}$
&25.81&5.73&3,830&28.97&5.11&4,683\\
\hspace{10pt}SPM
&34.12&9.30&4,863&32.14&9.57&6,590\\

\bottomrule
\end{tabular*}

{\scriptsize
\begin{minipage}{160mm}
Notes: The variables represent annual average concentration (ppb).
\end{minipage}
}

\end{center}
\end{table}
%------------------------------------

\subsection{Birth outcomes and infant mortality}\label{sec:appb2}

%------------------------------------
%Table 
\begin{table}[!t]
\begin{center}
\caption{Summary statistics of the birth and infant health outcomes}
\label{tab:sumhe}
\footnotesize

\begin{tabular*}{160mm}{l@{\extracolsep{\fill}}rrrrrr}
\toprule
&\multicolumn{3}{c}{Pre-treatment}&\multicolumn{3}{c}{Post-treatment}\\
\cmidrule(rl){2-4}
\cmidrule(rl){5-7}
&\multicolumn{1}{c}{Mean}&\multicolumn{1}{c}{SD}&\multicolumn{1}{c}{N}&\multicolumn{1}{c}{Mean}&\multicolumn{1}{c}{SD}&\multicolumn{1}{c}{N}\\
\hline

\textit{Regulated areas}
&&&&&&\\
\hspace{10pt}FDR
&37.24&11.38&260&34.82&11.45&260\\
\hspace{10pt}LBWR
&6.62&1.10&260&6.83&1.19&260\\
\hspace{10pt}IMR
&4.34&2.73&260&4.26&2.62&260\\
\hspace{10pt}NMR
&2.27&1.96&260&2.34&2.05&260\\
\\
\textit{Non-regulated areas}
&&&&&&\\
\hspace{10pt}FDR
&31.39&18.20&175&32.26&17.96&175\\
\hspace{10pt}LBWR
&6.10&2.68&175&6.98&3.07&175\\
\hspace{10pt}IMR
&4.23&6.39&175&4.22&6.71&175\\
\hspace{10pt}NMR
&2.13&4.49&175&2.26&5.09&175\\

\bottomrule
\end{tabular*}

{\scriptsize
\begin{minipage}{160mm}
Notes: FDR, LBWR, IMR, and NMR are the fetal death, low-birth weight, infant mortality, and neonatal mortality rates, respectively.
\end{minipage}
}

\end{center}
\end{table}
%------------------------------------

We used the FDR, LBWR, IMR, and NMR as our health outcome variables in Section~\ref{sec:health}.
The data on the number of fetal deaths, infant deaths, neonatal deaths, and births with a weight under 2,500 grams are taken from the 1991 and 1993 editions of the Vital Statistics of Japan (\textit{Jink\=od\=otai t\=okei}) published by the Statistics and Information Department, Minister's Secretariat, Ministry of Health and Welfare.
Several municipalities in a few islands in Tokyo prefecture are excluded because these islands are far from the main Japanese archipelago.
To deal with potential outliers, we trimmed the top 1\% municipalities that take greater values for each dependent variable.
Altogether, 435 municipalities in 1991 and 1993 are ultimately used in our dataset: 260 municipalities in regulated areas and 175 municipalities in non-regulated areas.
Table~\ref{tab:sumhe} shows the summary statistics of these health outcomes by area and period.
Most of these outcomes did not show a clear reduction in the post-treatment period, except for the FDR in regulated areas, consistent with our analytical result reported in Section~\ref{sec:health}.

\subsection{Control variables}\label{sec:appb3}

%------------------------------------
%Table 
\begin{table}[!t]
\begin{center}
\caption{Summary statistics of the control variables}
\label{tab:sumcov}
\footnotesize

\begin{tabular*}{160mm}{l@{\extracolsep{\fill}}ccc}
\toprule
&Mean&Standard Deviation&Observations\\
\hline

\textit{Meteorological variables}&&&\\
\hspace{10pt}Rainfall (days)	&117.06&26.50&20541\\
\hspace{10pt}Wind speed (days)	&15.98&23.88&17744\\
\hspace{10pt}Sunshine (\%)	&20.61&2.85&17290\\

\textit{Public health and income variables}
&&&\\
\hspace{10pt}Coverage of hospitals (‰)
&2.57&3.28&870\\
\hspace{10pt}Proportion of households receiving welfare benefits (\%)
&0.18&0.18&870\\\bottomrule
\end{tabular*}

{\scriptsize
\begin{minipage}{160mm}
Notes:
Rainfall is defined as the number of days with rainfall of over 1 mm.
Wind speed is defined as the number of days with a maximum wind speed of over 10 m/s.
Sunshine is defined as the percentage of sunshine hours in a year (8,760 hours).
Coverage of hospitals (‰) is defined as the number of hospitals per 1,000 households.
Proportion of poor households (\%) is defined as the number of households receiving welfare benefits per 100 households.
\end{minipage}
}

\end{center}
\end{table}
%------------------------------------

The meteorological data used are taken from the database of the Japan Meteorological Agency (\url{https://www.data.jma.go.jp/gmd/risk/obsdl/index.php}, accessed July 12, 2019).
These data consist of observations at more than 1,000 weather monitoring stations.
The data on the number of hospitals are obtained from the 1993 and 1995 editions of the \textit{Chiikiiry\=o kisot\=okei} (Statistics of the regional medical service) published by the Health, Labour and Welfare Statistics Association.
The data on the number of households receiving welfare benefits (\textit{Seikatsuhogo}) are taken from the 12 volumes of the Prefectural Statistics of Saitama, Chiba, Kanagawa, Tokyo, Osaka, and Kobe.
The details of these documents are listed in the references.
The data on the number of households are taken from the database of the Japanese Government Statistics (\url{https://www.e-stat.go.jp/stat-search/database?page=1&toukei=00200521&tstat=000000000023}, accessed November 1, 2019).
Coverage of hospitals (‰) is defined as the number of hospitals per 1,000 households.
This variable is measured at the municipal level.
The proportion of households receiving welfare benefits (\%) is defined as the number of households receiving benefits per 100 households.
Since a few prefectures were aggregated into this variable at the city-county-level, we used the average value of the proportion for those cases.
Table~\ref{tab:sumcov} presents the summary statistics of the control variables.

\section{Empirical analysis appendix}\label{sec:appc}
\setcounter{figure}{0} \renewcommand{\thefigure}{C.\arabic{figure}}
\setcounter{table}{0} \renewcommand{\thetable}{C.\arabic{table}}

\subsection{Air pollutant concentrations}\label{sec:appc1}
%------------------------------------
%Table 
\begin{table}
\begin{center}
\caption{Effects of the Automobile NO$_\text{x}$ Law on air pollutants}
\label{tab:result2}
\footnotesize

\begin{tabular*}{160mm}{l@{\extracolsep{\fill}}cccc}
\toprule

&(1) NO$_\text{x}$&(2) SO$_\text{2}$&(3) O$_\text{x}$&(4) SPM\\\hline

\textit{Regulation} $\times$ \textit{Post}
&$-$3.388***&$-$0.893***&$-$1.098***&$-$1.480***\\
&(0.525)&(0.101)&(0.224)&(0.358)\\\hline

Meteorological controls	&Yes&Yes&Yes&Yes\\
Station fixed effects	&Yes&Yes&Yes&Yes\\
Year fixed effects		&Yes&Yes&Yes&Yes\\

Observations
&14,085&13,743&9,286&12,193\\
Clusters
&\, 1,312&\, 1,274&\, \ 860&\, 1,186\\
Adj. $R$-squared&0.4516&0.5425&0.4322&0.3535\\
\bottomrule
\end{tabular*}

{\scriptsize
\begin{minipage}{160mm}
Notes: Each dependent variable represents annual average concentration (ppb). 
***, **, and * represent statistical significance at the 1\%, 5\%, and 10\% levels, respectively.
Standard errors clustered at the monitoring station level are in parentheses.
\end{minipage}
}
\end{center}
\end{table}
%------------------------------------

Table~\ref{tab:result2} reports the estimation results from the specification excluding the interaction term $\textit{Neighborhood}_{i} \times \textit{Post}_{t}$ in Eq.~\ref{eq:base}.
As shown in Columns~(1)--(4), the estimated effects of the regulation are statistically significantly negative across all specifications.
The estimates reported in Columns~(1)--(4) are close to those in Table~\ref{tab:result}.

\subsection{Birth and infant health outcomes}\label{sec:appc2}

%------------------------------------
%Table 
\begin{table}
\begin{center}
\caption{Effects of the Automobile NO$_\text{x}$ Law on fetal and infant health outcomes: Clustering the standard errors at the prefecture level}
\label{tab:didrob}
\footnotesize

\begin{tabular*}{160mm}{l@{\extracolsep{\fill}}cccc}
\toprule

		&(1) FDR	&(2) LBWR	&(3) IMR	&(4) NMR	\\
		&[37.24]	&[6.62]		&[4.34]		&[2.27]	\\\hline
\textit{Regulation}
&1.572&0.180&1.646&0.886\\
&(3.766)&(0.367)&(1.727)&(0.817)\\
\textit{Post}
&0.347&0.626**&$-$0.010&$-$0.003\\
&(0.764)&(0.111)&(0.467)&(0.138)\\
\textit{Regulation} $\times$ \textit{Post}
&$-$3.534**&$-$0.287&0.192&0.156\\
&(1.125)&(0.157)&(0.562)&(0.250)\\\hline

Control variables
&Yes&Yes&Yes&Yes\\
City-county FE
&Yes&Yes&Yes&Yes\\
Observations
&870&870&870&870\\
Municipalities
&435&435&435&435\\
Adj. $R$-squared
&0.4352&0.1777&0.0338&0.0191\\
\bottomrule
\end{tabular*}

{\scriptsize
\begin{minipage}{160mm}
Notes: 
FDR, LBWR, IMR, and NMR represent the fetal death rate, low-birth weight rate, infant mortality rate, and neonatal mortality rate, respectively.
The regression shown in Column (1) is weighted by the number of births, whereas the regressions shown in Columns (2)--(4) are weighted by the number of live births.
The mean values of the outcome variables in regulated areas in the pre-intervention period are reported in brackets.
The control variables include the coverage of hospitals and proportion of households receiving welfare benefits.
***, **, and * represent statistical significance at the 1\%, 5\%, and 10\% levels, respectively.
Standard errors clustered at the prefecture level are in parentheses.
\end{minipage}
}
\end{center}
\end{table}
%------------------------------------

In Section~\ref{sec:health}, we clustered standard errors at the city-county level to deal with the potential correlations among the residuals in cities and counties.
Thus, we assumed that all cities and counties are randomly sampled.
To check the sensitivity of this assumption, we further clustered standard errors at the prefecture level.
Table~\ref{tab:didrob} presents the results.
Clearly, the results are unchanged.
This finding supports the evidence that the potential correlations among cities and counties are negligible in our regressions.

%-------------------------------------------------------------------------------
% references
%-------------------------------------------------------------------------------

%-------------------------------------------------------------------------------
\end{document}